\documentclass{article}
\usepackage{spconf,amsmath,epsfig}
\usepackage{amssymb, bbm}
\usepackage[]{algorithm2e}
\usepackage{tabularx}
\usepackage{multirow, multicol}
\DeclareMathOperator*{\argmin}{argmin}
\usepackage{float}
\usepackage{multirow}		    
\usepackage{multicol}

\pdfoutput=1
\def\x{{\mathbf x}}
\def\vx{{\mathbf x}}
\def\vw{{\mathbf w}}

\newcommand{\norm}[1]{\left\lVert#1\right\rVert}

\title{ONLINE FEATURE RANKING FOR INTRUSION DETECTION SYSTEMS}
\name{Buse Gul Atli, Alexander Jung}
\address{Department of Computer Science, Aalto University, Finland; firstname.lastname@aalto.fi}
\begin{document}
%
\maketitle
\begin{abstract}
Many current approaches to the design of intrusion detection systems apply feature selection in a static, 
non-adaptive fashion. These methods often neglect the dynamic nature of network data which requires to 
use adaptive feature selection techniques. In this paper, we present a simple technique based on incremental 
learning of support vector machines in order to rank the features in real time within a streaming model for network data. 
Some illustrative numerical experiments with two popular benchmark datasets show that our approach allows to 
adapt to changing network behaviour and novel attack patterns.
\end{abstract}

\section{Introduction}
\label{sec:intro}
The design of efficient intrusion detection systems (IDS) has received considerable attention recently. 
An IDS aims at identifying and mitigating malicious activities. Most of 
the current IDSs can be divided into two main categories: signature-based and anomaly-based IDS \cite{Aaltodoc}. 

A signature-based IDS tries to detect intrusions by comparing the incoming network traffic to already 
known attacks, which are stored in the database as signatures. This class of IDS performs well in identifying 
known attacks but often fail to detect novel (unseen) attacks \cite{jacobus2015network}. 

The second category is referred to as anomaly-based IDS. Those IDS models the normal traffic
by learning patterns in the training phase. They label deviations from these learned patterns as
anomaly or intrusion \cite{denning1987intrusion}. 
The implementation of real-time anomaly-based IDS is challenging due to rapidly evolving network traffic behavior and limited 
amount of computational resources (computation time and memory) \cite{hu2014online}. 
Another challenge is the risk of overfitting due to the high-dimensional feature space and the
model complexity of IDS.
\cite{shrivas2014ensemble,sindhu2012decision,wang2010new,garcia2009anomaly}. 



In this paper we will address some of the short comings of existing anomaly-based IDS and present an efficient online feature ranking method. 
Our approach is based on online (or incremental) training of a support vector machine (SVM) which allows to cope with limited computational resources. 
The closest to our wok is \cite{hamed2018network, xia2009incremental}, which applies SVM for feature selection in a static setting. 
In contrast, our feature ranker finds approximate to the best feature subset in real-time and in the case of streaming network 
traffic with changing behaviour. Some illustrative numerical experiments indicate that real-time intrusion detection mechanisms 
using our feature ranker can be trained faster and achieves less error rate compared to offline
(batch) feature selection methods.

This paper is organized as follows: The problem setup is formalized in Section \ref{sec:problem}. In Section \ref{sec:Arch}, we detail 
the proposed online feature ranking method. The results of some illustrative numerical experiments are discussed in Section \ref{sec:exp}. 

\section{Problem Setup}
\label{sec:problem}

We consider a communication network within which information is exchanged using atomic units of data which are referred to as ``packets''. 
These packets consist of a header for control information and the actual payload (user data). 
We represent a packet by a feature vector $\x = (x_{1}, \cdots,x_{d}) \in \mathbb{R}^{d}$ with $d$ individual 
features $x_i$ listed in Table \ref{tab:Extracted features of the network traffic}. 
The last feature $x_{d}$ is a dummy feature being constant $x_{d}=1$. 

A machine learning based IDS aims at classifying network packets into the classes $\mathcal{Y} = \{ 0,1\}$. 
Each packet is associated with a binary label $y \in \mathcal{Y}$ with $y=1$ indicating a malicious packet (an ``attack'') and 
$y=0$ for a regular (normal) data packet. 
We view an IDS as a classifier $h: \mathbb{R}^{d} \rightarrow \mathcal{Y}$ which maps a given packet with feature vector $\x$, to a 
predicted label $\hat{y} = h(\x) \in \mathcal{Y}$ \cite{AGentleIntroML}. Naturally, the predicted label $\hat{y}$ should resemble the 
true class label $y$ as accurate as possible. Thus, we construct the classifier $h$ such that $y \approx h(\vx)$ for any packet with 
feature vector $\vx$ and true label $y$. In particular, we measure the classification error incurred when classifying a packet with 
feature $\vx$ and true label $y$ using the classifier $h$ by some loss function $\mathcal{L}((\vx,y),h)$. We will focus on a particular 
choice for the loss function as discussed in Section \ref{ssec:support}.

We learn the classifier $h$ in a supervised fashion by relying on a set of $N$ labeled packets $\{\x_i, y_i\}_{i=1}^{N}$ (the training set). 
In particular, we choose the classifier $h$ by minimizing the empirical risk 
\begin{equation} 
(1/N) \sum_{i=1}^{N} \mathcal{L}((\vx_{i},y_{i}),h) 
\end{equation}
incurred by $h$ on the training data set $\{ \vx_{i},y_{i}\}_{i=1}^{N}$. We will restrict ourselves to linear classifiers $h$, i.e., which have 
a linear decision boundary. The corresponding hypothesis space, constituted by all such linear classifiers, is
\begin{equation}
\label{equ_hypospace_lin_class}
\mathcal{H} = \{ h^{(\vw)}(\mathbf{x}) = \mathcal{I}( \mathbf{w}^{T} \vx \geq 0) \mbox{ for some } \vw \in \mathbb{R}^{d} \}. 
\end{equation} 
Here, we used the indicator function $\mathcal{I}(\mbox{``statement''}) \in \{0,1\}$ which is equal to one of the argument is a 
true statement and equal to zero otherwise. Using the parameterization \eqref{equ_hypospace_lin_class}, 
learning the optimal classifier for an IDS amounts to finding the optimal weight vector $\vw \in \mathbb{R}^{d}$: 
\begin{equation}
\label{equ_training_problem_weightvector}
\min_{\vw \in \mathbb{R}^{d}}  (1/N) \sum_{i=1}^{N} \mathcal{L}((\vx_{i},y_{i}),h^{(\vw)}). 
\end{equation}

It turns out that most of the relevant information for classifying network packets is often contained in a relatively small number of features due to redundant or irrelevant 
features \cite{guyon2003introductionVariable}. 
Thus, it is beneficial to perform some form of feature selection in order to avoid overfitting and improve the resulting 
classification performance. Moreover, for an online or real-time IDS, feature selection is instrumental in order to cope 
with a limited budget of computational resources. 

Most existing feature selection methods are based on statistical measures for dependency, such as Fisher score or (conditional) 
mutual information, and therefore independent of any particular classification method \cite{FScoreFeatSel,StructureFeatSelEventLogs}. 
In contrast, feature ranking methods use the entries $w_{j}$ fo the weight vector $\vw=(w_{1},\ldots,w_{j})$ of a particular classifier, e.g., 
obtained via solving \eqref{equ_training_problem_weightvector}, as a measure for the relevance of an individual feature $w_{j}$ \cite{chang2008feature}. 
We will apply feature ranking based on the weight vector obtained as the solution of \eqref{equ_training_problem_weightvector} 
for the particular loss function which is underlying the support vector machine (SVM). 


\section{General Architecture of The Feature Ranker}
\label{sec:Arch}
\begin{figure}[htb]
\begin{minipage}[b]{1.0\linewidth}
  \centering
  \centerline{\epsfig{figure=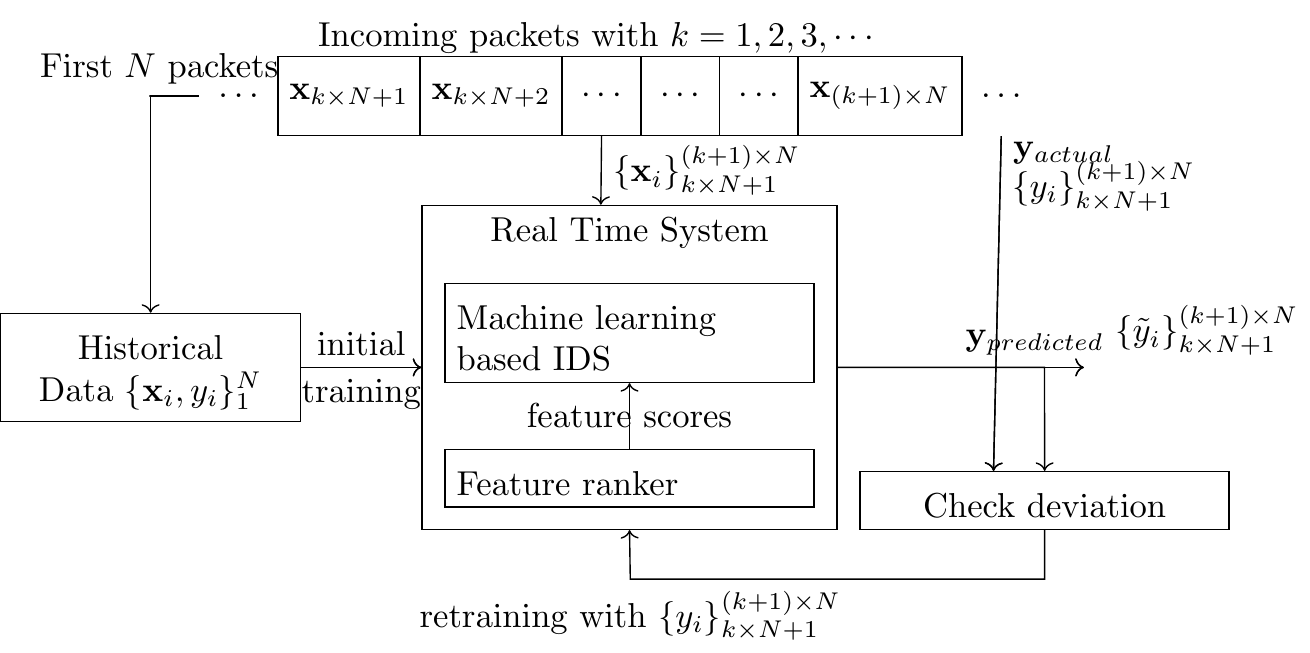,width=8.5cm}}
\end{minipage}
  \caption{General architecture of the proposed mechanism}
  \label{fig:General architecture of the proposed mechanism}
\end{figure}

Figure \ref{fig:General architecture of the proposed mechanism} shows the architecture of our system with an ability to rank 
features of the streaming data in real time based on the weights of the support vector machines (SVM) classifier. 
For the simplicity of our experiments, we used the same SVM also for the IDS part, but it might be any other intrusion detection mechanism using machine learning techniques.

Before training the feature ranker and the IDS, the first $N$ packets are collected as a historical data. These packets are used for 
pretraining the feature ranker and the IDS.  In our feature ranker, features critical to detect attacks have negative values, while 
features contributing for the normal traffic have positive values. After waiting another $N$ packets and predicting the 
label of each packet as normal or attack, the mean squared error
\begin{equation}
\frac{1}{N}\sum_{i=1}^{N}(y_i - \tilde{y}_i)^2
\end{equation}
is measured between the actual labels $y_i $ and the predicted labels $\tilde{y}_i$. If this error is quite high, then the feature 
ranker and the learning model is retrained by feeding the actual labels to the mechanism. Thus, feature weights are adapted 
to changes in normal network behaviour or new attack strategies.

\begin{table*}[htb]
\centering
\scalebox{0.70}{
\begin{tabular}{|c|c|c|c|c|c|}
\hline
 \bf No & \bf Feature & \bf No & \bf Feature & \bf No &\bf Feature\\ \hline
1 & ethernet type &  23 & IPv4 destination (divided into 4 features) & 35 & connection starting time\\ \hline
2 & sender MAC address (divided into 6 features) & 27 & TCP source port & 36 & fragmentation \\ \hline
8 & receiver MAC address ( divided into 6 features) & 28 & TCP destination port & 37 & overlapping\\ \hline
14 & IPv4 header length & 29 & UDP source port & 38 & ACK packet\\ \hline
15 & IP type of service & 30 & UDP destination port & 39 & retransmission \\ \hline
16 & IPv4 length & 31 & UDP/TCP length & 40 & pushed packet \\ \hline
17 & Time to live & 32 & ICMP type & 41 & SYN packet \\ \hline 
18 & IP protocol & 33 & ICMP code & 42 & FIN packet \\ \hline
19 &  IPv4 source (divided into 4 features) & 34 & duration of the connection & 43 & URG packet \\ \hline
\end{tabular}}
\caption{All features of a network packet used in online feature ranking mechanism.}

\label{tab:Extracted features of the network traffic}
\end{table*}
\subsection{Support Vector Machines (SVM)}
\label{ssec:support}

Our design of an IDS is based on \eqref{equ_training_problem_weightvector} using the particular loss 
loss function 
\begin{equation}\label{equ_reghingeloss}
\mathcal{L}((\vx,y),h) = \big[1- y (\vw^{T} \vx)\big]_{+} + (\lambda/2) \norm{\mathbf{w}}^2,
\end{equation}
with the shorthand  $[z]_{+} = \max \{0,z\}$. The loss function \eqref{equ_reghingeloss} consists of two components: 
The first component in \eqref{equ_reghingeloss} is known as the hinge loss and underlying the SVM classifier. 
The second component in \eqref{equ_reghingeloss} amounts to regularize the resulting classifier in order to avoid 
overfitting to the training data.   and a regularization parameter $\lambda > 0$ which allows 
to avoid overfitting. 
Inserting \eqref{equ_reghingeloss} into \eqref{equ_training_problem_weightvector} yields the weight of the SVM classifier  
\begin{equation} 
\label{equ_SVM_class}
\hat{\vw}\!=\!\argmin_{\vw \in \mathbb{R}^{d}} (1/N) \sum_{i=1}^{N} \big[1- y_{i} (\vw^{T} \x_{i})\big]_{+} + (\lambda/2) \norm{\mathbf{w}}^2. 
\end{equation} 
Note that \eqref{equ_SVM_class} is a non-smooth convex optimization problem which precludes the application 
of basic gradient based optimization methods \cite{JungFixedPoint}. Instead, as detailed in Section \ref{ssec:stoch}, we apply 
the stochastic sub-gradient method to solve \eqref{equ_SVM_class} in an online fashion. 

Once we have learnt the optimal weight vector $\hat{\vw} = (\hat{w}_{1},\ldots,\hat{w}_{d})^{T}$ by solving \eqref{equ_SVM_class}, we can classify newly arriving 
packet with feature $\vx$.
Therefore, for any testing sample $\x_i$, the SVM classifier $h(\cdot)$ can be written as
\begin{equation}
h(\vx) = \mathcal{I} (\hat{\mathbf{w}}^T\vx > 0 ). 
\end{equation}
Thus, the classification result depends on the features $x_{i}$ via the weights $\hat{w}_{i}$. A weight $\hat{w}_{i}$ having large 
absolute value implies that the feature $x_{i}$ has a strong influence no the classification result \cite{guyon2003introduction}. 
It is therefore reasonable to keep only those features $x_{i}$ whose weights $w_{i}$ are largest and discard the 
remaining features in order to save computational resources and avoid overfitting. 



\subsection{Stochastic Gradient Descent (SGD)}
\label{ssec:stoch}

A naive implementation of SVM does not scale well due to excessive computational 
time and memory requirements \cite{zheng2013online}. In order to cope with the requirements of real-time IDS one typically needs to use online implementations 
of SVM. A simple modification of the linear SVM method 
can be done by incorporating gradient methods and adapting optimum 
weight vector as new data arrives.

Since SVM is trying to solve a convex cost function, we can find a solution by using iterative optimization algorithms. 
Stochastic gradient descent is one of the optimization algorithm that minimizes the cost function by performing the 
gradient descent on a single or few examples. in SGD, each $k$-th iteration updates the weights of the learning 
algorithms including SVM, percepton or linear regression, with respect to the gradient of the loss function $\mathcal{L}$

\begin{equation}
\mathbf{w}^{k+1} = \mathbf{w}^{k} -  \alpha \nabla_{\mathbf{w}}\mathcal{L}_n(h(x),y),
\end{equation} 
where $\alpha$ denotes constant learning rate and $n$ is a set of indices randomly sampled from the training set $\{\x_i, y_i\}_{i=1}^{N}$. In our setup, we approximate the gradient by 
drawing $M$ points uniformly over the training data and perform mini batch iterations of SGD over SVM. 
Therefore the updates of weight vector $\mathbf{w}$ will become

\begin{equation}\label{gradupdate}
\mathbf{w}^{k+1} = \mathbf{w}^{k} - \alpha \frac{1}{M}\sum_{j=0}^{M}\nabla_{\mathbf{w}}\mathcal{L}(g(\x)_j, y_j)
\end{equation}

Although the hinge loss is not differentiable, we can still calculate the subgradient with 
respect to the weight vector $\mathbf{w}$. The derivative of the hinge loss can be written as:
\begin{equation}
\frac{\partial \big[1- y_{i} (\vw^{T} \x_{i})\big]_{+}}{\partial w_k} = \left\{
        \begin{array}{ll}
            -y_ix_i & \quad y_i(\mathbf{w}_{k}^T \x_{i} + b_{i}) < 1 \\
            0 & \text{otherwise}.
        \end{array}
    \right.
\end{equation}
Therefore, the gradient update for Equation \ref{gradupdate} at $k$-th iteration is calculated as:
\begin{equation}
\mathbf{w}^{k+1} = \mathbf{w}^{k} - \alpha \lambda \mathbf{w}_k - \frac{1}{M}\sum_{j=0}^{M}\mathbbm{1}[\, (\mathbf{w}_k^T\x_{j} + b_{j}) < 1  ]\,y_{j}\x_j,
\end{equation}
where $\mathbbm{1}[\, (\mathbf{w}_k^T\x_{j} + b_{j}) < 1  ]\, $ is an indicator function which results in one when its argument is true, and zero otherwise. 

In large scale learning problems, SGD decreases the time complexity, since it does not need 
all the data points to calculate the loss. 
Although, it only samples few data point to reach the optimum point, it provides a reasonable solution to the optimization problem in the case of big data \cite{bottou2010large}. 

In our proposed feature ranker mechanism, the weight vector of a SVM classifier is optimized in an online fashion using SGD with mini batch size $N$. After predicting the labels of packets in the next mini-batch, the predicted labels are compared to the true classes of packets, which are decided by experts. If the predicted labels highly deviate from the true labels based on the MSE, then the weights are updated according to this $N$ packets. This also enables an online correction of the feature weights, since the weight vector is updated based on the incoming $N$ packets with a possible different connections, applications or attacks. 

\section{Experimental Results}
\label{sec:exp}

We verified the performance of the proposed IDS by means of numerical experiements run on a 64bit Ubuntu 16.04 PC with 8 GiB of 
RAM and CPU of 2.70GHz. Two different datasets, the ISCX IDS 2012 dataset and CIC Android Adware-General Malware, were used 
for our experiments to test both batch and online algorithms. The ISCX-IDS 2012 dataset \cite{shiravi2012toward} was generated by the 
Canadian Institute for Cybersecurity and mostly used for anomaly based intrusion detection systems. This dataset includes seven days 
of traffic and produces the complete capture of the all network traces with HTTP, SMTP, SSH, IMAP, POP3, and FTP protocols. It contains 
multi-stage intrusions such as SQL injection, HTTP DoS, DDoS and SSH attacks. The CIC Android Adware-General Malware 
dataset \cite{lashkaritowards} is also produced by the same research group. This dataset consists of malwares, including viruses, 
worms, trojans and bots, on over 1900 Android applications.

Both datasets are provided as pcap files including both malicious network packets and normal packets. We selected different malware 
types (InfoStl, Dishigy, Zbot, Gamarue) for testing batch (offline) feature ranking. We used the ISCX-IDS 2012 dataset for the online feature 
ranking, since it has the complete capture of one week network traffic data without any interruption to the overall connection. From each dataset, we 
transferred raw packets to a file with 43 features using ``tshark'' network analyzer. Table \ref{tab:Extracted features of the network traffic} lists all the features measured with tshark and used in the experiments.

The first set of experiments does not include online incremental learning of SVM weights. It implements the linear SVM 
classifier with SGD update, as well as other widely used feature selection algorithms, selects 20 features with the highest weight out of 43, 
and uses them as an input to a single-layer feedforward neural network with 10 hidden neurons and sigmoid activation function. 
The experiments were performed on various pcap files from the CIC Android dataset having different sparsity, attacks and malwares in order to prove 
that feature ranking with linear SVM weights in batch (offline) learning provides good performance compared to the 
other feature selection methods. Table \ref{tab:Comparison of offline feature selection methods} shows the experimental 
results of these feature selection methods by measuring the time spent on extracting the best features from the training 
set and the accuracy of the classification. It can be seen that linear SVM weights trained with the SGD method is quite 
fast and generally gives a higher accuracy than the other methods. Therefore, we can suggest that incremental version 
of the linear SVM approach can be extended to the case of streaming network traffic.

\begin{table}[htb]
\centering
\scalebox{0.70}{
\begin{tabular}{c cc cc cc}
\multirow{2}{*}{Malware type} & \multicolumn{2}{c}{F-score} & \multicolumn{2}{c}{Decision Trees} & \multicolumn{2}{c}{Chi2} \\ 
 & time (sec.) & accuracy & time & accuracy & time & accuracy \\ \hline
InfoStl          &  \bf 0.005  & 0.891 & 0.022 & 0.993 & \bf 0.005 & 0.992 \\
Dishigy         &  0.002  & 0.651 & 0.018 & 0.987 &  0.002 & \bf 1.0 \\
Zbot             &  0.002  & 0.970 & 0.017 & 0.989 & 0.002 & 0.987\\
Gamarue       &  0.002  & 0.933 & 0.016 & 0.947 & 0.002 & 0.947 \\
 & \multicolumn{2}{c}{Mutual info.} & \multicolumn{2}{c}{RFE} & \multicolumn{2}{c}{SVM weights} \\
 & time (sec.) & accuracy & time & accuracy & time & accuracy \\ \cline{2-7}
 InfoStl    &  2.503& 0.993  & 1.208  & 0.995 & \bf 0.005 & \bf 0.995 \\
 Dishigy   & 0.364 &  1.0     & 0.121 & 1.0      & \bf 0.001 &\bf 1.0  \\
 Zbot       & 0.633 &  0.988 & 0.241 & 0.987 &  \bf 0.001 & \bf 0.989\\
 Gamarue & 0.240 &  0.946 & 0.088 & \bf 0.969 &   \bf 0.001 &0.964 \\ \hline
 \multicolumn{6}{l}{\bf Bold:best result achieved} \\ \hline
\end{tabular}}
\caption{Feature selection comparison with different malwares}
\label{tab:Comparison of offline feature selection methods}
\end{table}

\begin{figure}[htb]
\begin{minipage}[b]{1.0\linewidth}
  \centering
  \centerline{\epsfig{figure=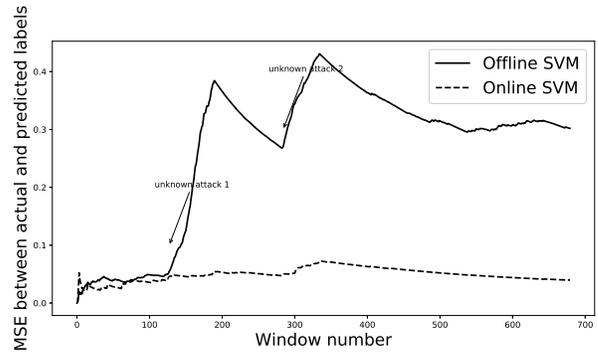,width=8cm}}
\end{minipage}
\caption{MSE of the streaming test data with a load change}
  \label{fig:MSE of the streaming test data with a load change}
\end{figure}

The second experiment was implemented with both offline and online SVM with a linear kernel in the case of streaming network data. Both feature weighting and classification is done by linear SVM in either online or offline fashion. Experiments were performed with the ISCX-IDS 2012 
dataset. Figure \ref{fig:MSE of the streaming test data with a load change} shows the mean squared error between the true and predicted 
labels of the streaming data while the network behaviour changes and new type of attacks are fed to the input. Unlike the offline SVM, online SVM changes its weight vector in a sliding window with $N$ packets.The training was implemented in the first $100*N$ packets which spans one day of network activities including injection attacks. 
The second day comes with a more complex and harder to detect injection attack and the third day also starts with a DoS attack. 
As can be seen in the figure, our online feature ranker model is able to adapt itself to a change in the network statistics. However, 
batch learning with SVM performs poorly when the arriving packets shows a completely different behaviour. In conclusion, on-line models 
quickly adapt to changing behaviour in the network data and achieves a notable improvement on the performance of prediction over batch learning.

\begin{figure}[htb]
\begin{minipage}[b]{1.0\linewidth}
  \centering
  \centerline{\epsfig{figure=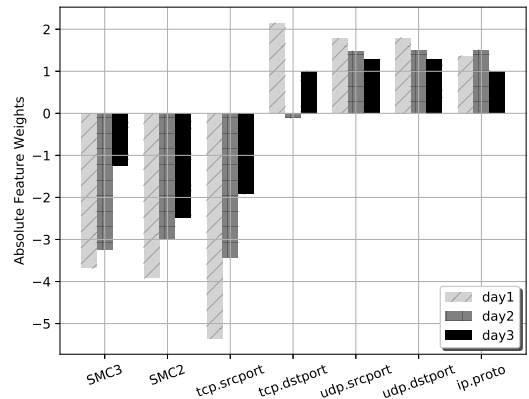,width=8cm}}
\end{minipage}
 \caption{Feature Weights for Different Network Behaviors}
 \label{fig:Feature Weights for Different Network Behaviors}
\end{figure}

The last experiment was also implemented with the ISCX-IDS 2012 dataset by choosing 3 days of network data with different attacks to 
prove that the importance of features are different for each attack. Figure \ref{fig:Feature Weights for Different Network Behaviors} shows 
the resulting absolute weights of randomly selected 5 features. Figure \ref{fig:Feature Weights for Different Network Behaviors} indicates 
that feature weights change even from positive to negative in each day having different connections and different attack types. These 
results prove that our feature ranker mechanism works as we desired and adjusts the feature weights with streaming network data.

\section{Conclusion}
\label{sec:conclusion}

In this paper, we proposed a simple method based on incremental learning of linear SVM models to rank or weight features in real time. 
Experimental results indicated that our method can adjust the importance of any feature based on not only the changing network behaviour 
but also for novel attacks. In future research, the online feature ranking with a variable chunk size will be incorporated to better fine-tuning 
of our proposed method. In addition, the weights of the features will be used as input layer weights of neural networks based IDSs, which 
can be used as real-time and self-trained intrusion detection mechanism. 




\begin{thebibliography}{10}

\bibitem{Aaltodoc}
Buse Atli,
\newblock ``Anomaly-based intrusion detection by modeling probability
  distributions of flow characteristics,''
\newblock M.S. thesis, Aalto University, Espoo, Finland, 2017-10-23.

\bibitem{jacobus2015network}
Agustinus Jacobus and Alicia~A.E. Sinsuw,
\newblock ``Network packet data online processing for intrusion detection
  system,''
\newblock in {\em 2015 1st International Conference on Wireless and Telematics
  (ICWT)}. IEEE, 2015, pp. 1--4.

\bibitem{denning1987intrusion}
Dorothy~E. Denning,
\newblock ``An intrusion-detection model,''
\newblock {\em IEEE Trans. Softw. Eng.}, vol. 13, no. 2, pp. 222--232, Feb.
  1987.

\bibitem{hu2014online}
Weiming Hu, Jun Gao, Yanguo Wang, Ou~Wu, and Stephen Maybank,
\newblock ``Online adaboost-based parameterized methods for dynamic distributed
  network intrusion detection,''
\newblock {\em IEEE Transactions on Cybernetics}, vol. 44, no. 1, pp. 66--82,
  2014.

\bibitem{shrivas2014ensemble}
Akhilesh~Kumar Shrivas and Amit~Kumar Dewangan,
\newblock ``An ensemble model for classification of attacks with feature
  selection based on kdd99 and nsl-kdd data set,''
\newblock {\em International Journal of Computer Applications}, vol. 99, no.
  15, 2014.

\bibitem{sindhu2012decision}
Siva S~Sivatha Sindhu, S~Geetha, and Arputharaj Kannan,
\newblock ``Decision tree based light weight intrusion detection using a
  wrapper approach,''
\newblock {\em Expert Systems with applications}, vol. 39, no. 1, pp. 129--141,
  2012.

\bibitem{wang2010new}
Gang Wang, Jinxing Hao, Jian Ma, and Lihua Huang,
\newblock ``A new approach to intrusion detection using artificial neural
  networks and fuzzy clustering,''
\newblock {\em Expert systems with applications}, vol. 37, no. 9, pp.
  6225--6232, 2010.

\bibitem{garcia2009anomaly}
Pedro Garcia-Teodoro, J~Diaz-Verdejo, Gabriel Maci{\'a}-Fern{\'a}ndez, and
  Enrique V{\'a}zquez,
\newblock ``Anomaly-based network intrusion detection: Techniques, systems and
  challenges,''
\newblock {\em computers \& security}, vol. 28, no. 1, pp. 18--28, 2009.

\bibitem{hamed2018network}
Tarfa Hamed, Rozita Dara, and Stefan~C Kremer,
\newblock ``Network intrusion detection system based on recursive feature
  addition and bigram technique,''
\newblock {\em Computers \& Security}, vol. 73, pp. 137--155, 2018.

\bibitem{xia2009incremental}
Yong-Xiang Xia, Zhi-Cai Shi, and Zhi-Hua Hu,
\newblock ``An incremental svm for intrusion detection based on key feature
  selection,''
\newblock in {\em Intelligent Information Technology Application, 2009. IITA
  2009. Third International Symposium on}. IEEE, 2009, vol.~3, pp. 205--208.

\bibitem{AGentleIntroML}
A.~Jung,
\newblock ``A {G}entle {I}ntroduction to {S}upervised {M}achine {L}earning,''
\newblock {\em arXiv}, 2018.

\bibitem{guyon2003introductionVariable}
Isabelle Guyon and Andr{\'e} Elisseeff,
\newblock ``An introduction to variable and feature selection,''
\newblock {\em Journal of machine learning research}, vol. 3, no. Mar, pp.
  1157--1182, 2003.

\bibitem{FScoreFeatSel}
S.~Ding,
\newblock ``Feature selection based f-score and aco algorithm in support vector
  machine,''
\newblock in {\em 2009 Second International Symposium on Knowledge Acquisition
  and Modeling}, Nov 2009, vol.~1, pp. 19--23.

\bibitem{StructureFeatSelEventLogs}
M.~Hinkka, T.~Lehto, K.~Heljanko, and A.~Jung,
\newblock ``Structural feature selection for event logs,''
\newblock in {\em Business Process Management Workshops. BPM 2017.}, 2017.

\bibitem{chang2008feature}
Yin-Wen Chang and Chih-Jen Lin,
\newblock ``Feature ranking using linear svm,''
\newblock in {\em Causation and Prediction Challenge}, 2008, pp. 53--64.

\bibitem{JungFixedPoint}
A.~Jung,
\newblock ``A fixed-point of view on gradient methods for big data,''
\newblock {\em Frontiers in Applied Mathematics and Statistics}, vol. 3, 2017.

\bibitem{guyon2003introduction}
Isabelle Guyon and Andr{\'e} Elisseeff,
\newblock ``An introduction to variable and feature selection,''
\newblock {\em Journal of machine learning research}, vol. 3, pp. 1157--1182,
  2003.

\bibitem{zheng2013online}
Jun Zheng, Furao Shen, Hongjun Fan, and Jinxi Zhao,
\newblock ``An online incremental learning support vector machine for
  large-scale data,''
\newblock {\em Neural Computing and Applications}, vol. 22, no. 5, pp.
  1023--1035, 2013.

\bibitem{bottou2010large}
L{\'e}on Bottou,
\newblock ``Large-scale machine learning with stochastic gradient descent,''
\newblock in {\em Proceedings of COMPSTAT'2010}, pp. 177--186. Springer, 2010.

\bibitem{shiravi2012toward}
Ali Shiravi, Hadi Shiravi, Mahbod Tavallaee, and Ali~A Ghorbani,
\newblock ``Toward developing a systematic approach to generate benchmark
  datasets for intrusion detection,''
\newblock {\em computers \& security}, vol. 31, no. 3, pp. 357--374, 2012.

\bibitem{lashkaritowards}
Arash~Habibi Lashkari, Andi Fitriah~A Kadir, Hugo Gonzalez, Kenneth~Fon Mbah,
  and Ali~A Ghorbani,
\newblock ``Towards a network-based framework for android malware detection and
  characterization,''
\newblock in {\em 15th International Conference on Privacy, Security and Trust
  (PST)}, 2017.

\end{thebibliography}

\end{document}